\pdfoutput=1

\RequirePackage{fix-cm}

\documentclass[twocolumn]{svjour3}

\smartqed

\usepackage{graphicx}
\usepackage{amssymb,amsmath}
\usepackage{txfonts}
\usepackage{natbib}
\usepackage{hyperref}

\hypersetup{pdftitle = The title of my PDF, pdfauthor = My name, pdfsubject= The subject, pdfkeywords = keyword1 keyword2 keyword3}
\hypersetup{colorlinks = true, linkcolor = blue, anchorcolor = red, citecolor = blue, filecolor = red, pagecolor = red, urlcolor = blue}

\journalname{Astrophysics \& Space Science}

\newcommand{\mathz}{\ooalign{$z$\cr\hfil\rule[.5ex]{.2em}{.06ex}\hfil\cr}}

\begin{document}

\title{Comparing the basins of attraction for several methods in the circular Sitnikov problem with spheroid primaries}

\author{Euaggelos E. Zotos}

\institute{Department of Physics, School of Science, \\
Aristotle University of Thessaloniki, \\
GR-541 24, Thessaloniki, Greece\\
Corresponding author's email: {evzotos@physics.auth.gr}}

\date{Received: 4 March 2018 / Accepted: 2 May 2018}

\titlerunning{Basins of attraction in the Sitnikov problem with spheroid primaries}

\authorrunning{Euaggelos E. Zotos}

\maketitle

\begin{abstract}

The circular Sitnikov problem, where the two primary bodies are prolate or oblate spheroids, is numerically investigated. In particular, the basins of convergence on the complex plane are revealed by using a large collection of numerical methods of several order. We consider four cases, regarding the value of the oblateness coefficient which determines the nature of the roots (attractors) of the system. For all cases we use the iterative schemes for performing a thorough and systematic classification of the nodes on the complex plane. The distribution of the iterations as well as the probability and their correlations with the corresponding basins of convergence are also discussed. Our numerical computations indicate that most of the iterative schemes provide relatively similar convergence structures on the complex plane. However, there are some numerical methods for which the corresponding basins of attraction are extremely complicated with highly fractal basin boundaries. Moreover, it is proved that the efficiency strongly varies between the numerical methods.


\end{abstract}

\section{Introduction}
\label{intro}

A special version of the classical restricted three-body problem is the so-called Sitnikov problem \citep{S60}. In this case, two equally massed primary bodies move in circular or elliptic orbits, while the test particle oscillates along the vertical $z$ axis, perpendicular to the orbital plane $(x,y)$ of the primaries. The simplest case, where the two primary bodies move in circular orbits is also known as the MacMillan problem \citep{McM11}. Over the years, a large number of studies have been devoted on the Sitnikov problem, while introducing various types of perturbation such as the radiation pressure \citep[e.g.,][]{PK06}, the prolateness of the primaries \citep[e.g.,][]{DKMP12}, as well as the oblateness of the primaries \citep[e.g.,][]{RGH15}.

In dynamical astronomy and celestial mechanics the equilibrium points of a system play a role of great importance since at these locations the test particle is able to maintain its relative position, with respect to the primary bodies. This is true because at the libration points of the system the combined gravitational attraction of the primaries provides precisely the required centripetal force. Unfortunately, in many systems, such as those of the $N$-body problem (with $N \geq 3$), there are no explicit formulae for the positions of the libration points. Therefore, the locations of the equilibrium points can be obtained only by means of numerical methods. In other words, we need a multivariate iterative scheme for solving the system of the first order derivatives. It is well known that the results of any numerical method strongly depend on the initial conditions (staring points of the iterative procedure). Indeed, for some initial conditions the iterative formulae converge quickly, while for other starting points a considerable amount of iterations is required for reaching to a root (equilibrium point). Fast converging points usually belong to basins of attraction, while on the other hand slow converging points are located in fractal regions. On this basis, the knowledge of the basins of attraction of a dynamical system is very important because these basins reveal the optimal (regarding fast convergence) starting points for which the iterative formulae require the lowest amount of iterations, for leading to an equilibrium point. In addition, being aware of the fractal regions we know exactly which points should be avoided as initial conditions of the iterative formulae.

Furthermore, the basins of attraction, associated with the equilibrium points, contain useful information regarding the intrinsic dynamical properties of a system. This should be true if we take into account that the iterative scheme usually contains both the first and the second order derivatives of the effective potential. It is well know that the first order derivatives are directly linked with the equations of motion of the test particle, while the second order derivatives are used for computing the variational equations. In addition, the variational equations are used for the calculation of the monodromy matrix of the periodic orbits and therefore they are also linked with the stability properties of the test particle. All the above-mentioned issues justify the reasons of why one should be aware of the basins of attraction in a dynamical system.

The literature is replete of papers on the basins of attraction in several types of dynamical systems. In the vast majority of them the Newton-Raphson iterative scheme (which is the simplest one) is used for revealing the convergence properties in several dynamical systems such as the Sitnikov problem \citep[e.g.,][]{DKMP12}, the Hill problem with oblateness and radiation pressure \citep[e.g.,][]{D10,Z17b}, the circular restricted three-body problem with oblateness and radiation pressure \citep[e.g.,][]{Z16}, the Copenhagen problem with radiation pressure \citep[e.g.,][]{K08}, the pseudo-Newtonian planar circular restricted three-body problem \citep[e.g.,][]{Z17c}, the circular restricted four-body problem \citep[e.g.,][]{BP11,KK14,Z17a,Z17d}, the circular restricted four-body problem with radiation pressure \citep[e.g.,][]{APHS16}, the circular restricted four-body problem with various perturbations \citep[e.g.,][]{SAA17,SAP17}, the circular restricted five-body problem \citep[e.g.,][]{ZS18}, the ring problem of $N + 1$ bodies \citep[e.g.,][]{CK07,GKK09}, or even the restricted 2+2 body problem \citep[e.g.,][]{CK13}.

In \citet{DKMP12} the Newton-Raphson basins of attraction of the Sitnikov problem with prolate primaries have been briefly investigated. In the present paper will use a large variety of numerical methods in an attempt to reveal the corresponding basins of convergence. In addition, we will try to evaluate each iterative scheme and therefore obtain a general overview regarding the convergence speed as well as the efficiency of all the numerical methods.

The structure of the present article is as follows: the basic properties of the dynamical model are presented in Section \ref{mod}. In Section \ref{eqpts} we discuss the parametric evolution of the roots of the system. The following Section contains all the numerical outcomes, regarding the basins of attraction of all the numerical methods. The paper ends with Section \ref{conc}, where we provide the main conclusions of our analysis.

\section{The mathematical model}
\label{mod}

The dynamical system is mainly composed of two primary bodies, $P_1$ and $P_2$, which move, around their common center of gravity, in circular orbits. The dimensionless masses of the primaries are $m_1 = \mu$ and $m_2 = 1 - \mu$, respectively, where $\mu = m_2/(m_1 + m_2) \leq 1/2$ is the mass parameter \citep{S67}. A third body (which behaves as a test particle) is moving in the combined gravitational field of the primaries however, its motion does not perturb the orbits of the two main bodies. This is because the mass of the third body $m$ is considerable smaller with respect to the masses of the primaries. In a dimensionless rotating system of coordinates $Oxyz$ the centers of the two main bodies are located at $(x_1, 0, 0)$ and $(x_2, 0, 0)$, where $x_1 = - \mu$ and $x_2 = 1 - \mu$.

It is assumed that the primary bodies do not have a spherically symmetric shape but they resemble a spheroid. For this reason we introduce the oblateness coefficient $A_i$, $i = 1,2$. When $A < 0$ the primary is a prolate spheroid, while when $A > 0$ the primary is an oblate spheroid.

According to \citet{AS06,DM06,OV03,SSR75} the time-independent effective potential of the circular restricted-three
body problem, where the primaries are spheroids is given by
\begin{equation}
\Omega(x,y,z) = \sum_{i=1}^{2} \frac{m_i}{r_i}\left(1 + \frac{A_i}{2r_i^2} - \frac{3A_i z^2}{2r_i^4}\right) + \frac{n^2}{2} \left(x^2 + y^2 \right),
\label{pot}
\end{equation}
where
\begin{align}
r_1 &= \sqrt{\left(x - x_1 \right)^2 + y^2 + z^2}, \nonumber\\
r_2 &= \sqrt{\left(x - x_2 \right)^2 + y^2 + z^2},
\label{dist}
\end{align}
are the distances of the test particle from the respective primary bodies, while $n$ is the mean motion of the primaries, which is defined as
\begin{equation}
n = \sqrt{1 + 3\left(A_1 + A_2 \right)/2}.
\label{mn}
\end{equation}

The equations which govern the motion of a test particle read
\begin{align}
&\ddot{x} - 2 n \dot{y} = \frac{\partial \Omega}{\partial x}, \nonumber\\
&\ddot{y} + 2 n \dot{x} = \frac{\partial \Omega}{\partial y}, \nonumber\\
&\ddot{z} = \frac{\partial \Omega}{\partial z}.
\label{eqmot}
\end{align}

This dynamical system admits only one integral of motion (also known as the Jacobi integral). The corresponding Hamiltonian is
\begin{equation}
J(x,y,z,\dot{x},\dot{y},\dot{z}) = 2\Omega(x,y,z) - \left(\dot{x}^2 + \dot{y}^2 + \dot{z}^2 \right) = C,
\label{ham}
\end{equation}
where of course $\dot{z}$, $\dot{y}$, and $\dot{z}$ are the velocities, while $C$ is the numerical value of the Jacobi constant which is conserved.

\begin{figure}[!t]
\centering
\resizebox{\hsize}{!}{\includegraphics{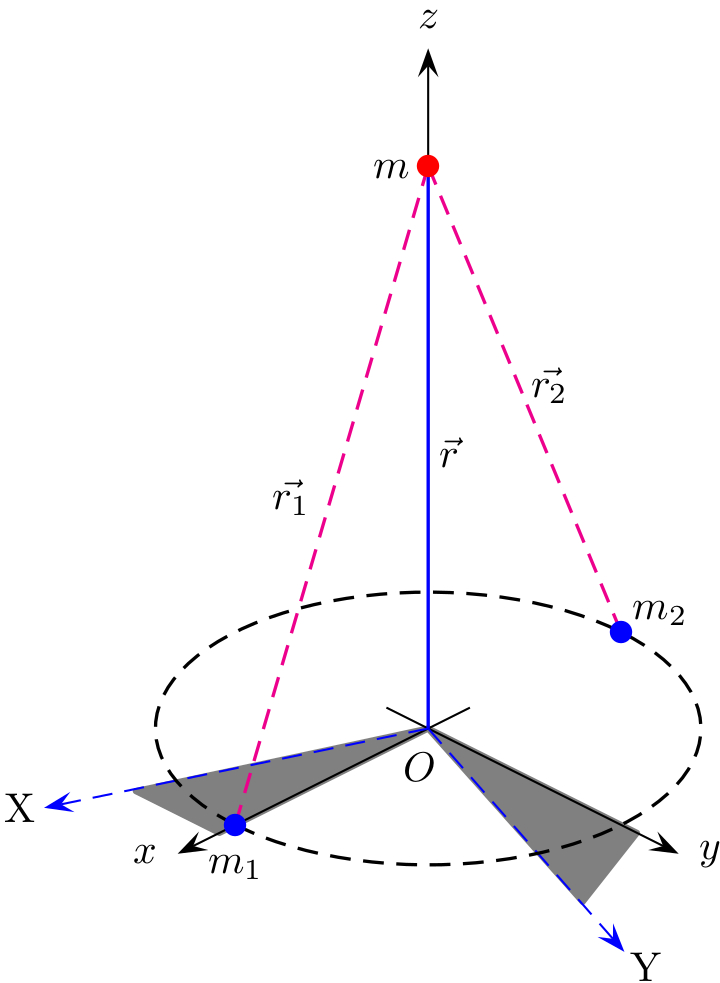}}
\caption{The spatial configuration of the circular Sitnikov problem, where the two equally massed primary bodies $(m_1 = m_2 = 1/2)$ move on symmetric circular orbits on the $(x,y)$ plane. The test particle, with mass $m$ oscillates in a straight line, perpendicular to the orbital plane of the primaries. (Color figure online).}
\label{conf}
\end{figure}

In order to obtain the potential function of the circular Sitnikov problem all we have to do is to set $\mu = 1/2$, $x = y = 0$, and $A_1 = A_2 = A$ in Eq. (\ref{pot}). Then we obtain
\begin{equation}
\Omega(z) = \frac{1}{r} + \frac{A}{2r^3} - \frac{3Az^2}{2r^5},
\label{potz}
\end{equation}
where $r = \sqrt{z^2 + 1/4}$. Looking at Eq. (\ref{potz}) we realize that it describes the motion of a massless test particle, oscillating along a straight line which is perpendicular to the orbital $(x,y)$ plane of the two equally massed primaries. The spatial geometry of circular Sitnikov problem is presented in Fig. \ref{conf}.

The motion of the test particle, along the vertical $z$ axis, is described by the equation
\begin{equation}
\ddot{z} = - \frac{z}{r^3} - \frac{9Az}{2r^5} + \frac{15Az^3}{2r^7},
\label{eqmotz}
\end{equation}
while the corresponding energy (Jacobi) integral, regarding the vertical motion, has the form
\begin{equation}
J(z,\dot{z}) = 2 \Omega(z) - \dot{z}^2 = C_{z}.
\label{hamz}
\end{equation}

\section{The equilibrium points (roots) of the system}
\label{eqpts}

Following the approach successfully used in \citet{DKMP12} (see Section 3), from now on the $z$ coordinate is considered as a complex variable and it is denoted by $\mathz$. The transition to complex numbers is imperative because all the impressive fractal basin structures appear only on the complex plane, as it was discussed in \citet{D10}.

In order to locate the positions of the equilibrium points (roots) we have to set the right hand side of Eq. (\ref{eqmotz}) equal to zero as
\begin{equation}
f(\mathz;A) = - \frac{8\mathz\left(16 \mathz^4 + 8 \left(1 - 6A \right)\mathz^2 + 18 A + 1 \right)}{\left(1 + 4\mathz^2\right)^{7/2}} = 0,
\label{fza0}
\end{equation}
which is reduced to
\begin{equation}
\mathz \left(16 \mathz^4 + 8 \left(1 - 6A \right)\mathz^2 + 18 A + 1 \right) = 0.
\label{fza}
\end{equation}

Looking at Eq. (\ref{fza}) we observe that the root $\mathz = 0$ is always present, regardless the value $A$ of the oblateness coefficient. This root corresponds to the inner collinear equilibrium point $L_1$ of the circular restricted three-body problem. However since the left hand side of Eq. (\ref{fza}) is a fifth order polynomial it means that there are four additional roots, given by
\begin{equation}
\mathz_i = \pm \frac{1}{2} \sqrt{6A - 1 \pm \sqrt{6A \left(6A - 5\right)}}, \ \ \ i = 1,...,4.
\label{rts}
\end{equation}

The nature of these four roots strongly depends on the numerical value $A$ of the oblateness coefficient. Our analysis reveals that, along with the $\mathz = 0$ root
\begin{itemize}
  \item When $A < -1/18$ there are two real and two imaginary roots.
  \item When $A = -1/18$ there are two imaginary roots.
  \item When $A \in (-1/18,0)$ there are four imaginary roots.
  \item When $A = 0$ only the root $\mathz = 0$ exists.
  \item When $A \in (0, 5/6)$ there are four complex roots.
  \item When $A = 5/6$ there are two real roots.
  \item When $A > 5/6$ there are four real roots.
\end{itemize}
It is seen, that the values $A = \{-1/18, 0, 5/6 \}$ are in fact critical values of the oblateness coefficient, since they determine the change on the nature of the four roots.

\begin{figure}
\centering
\resizebox{\hsize}{!}{\includegraphics{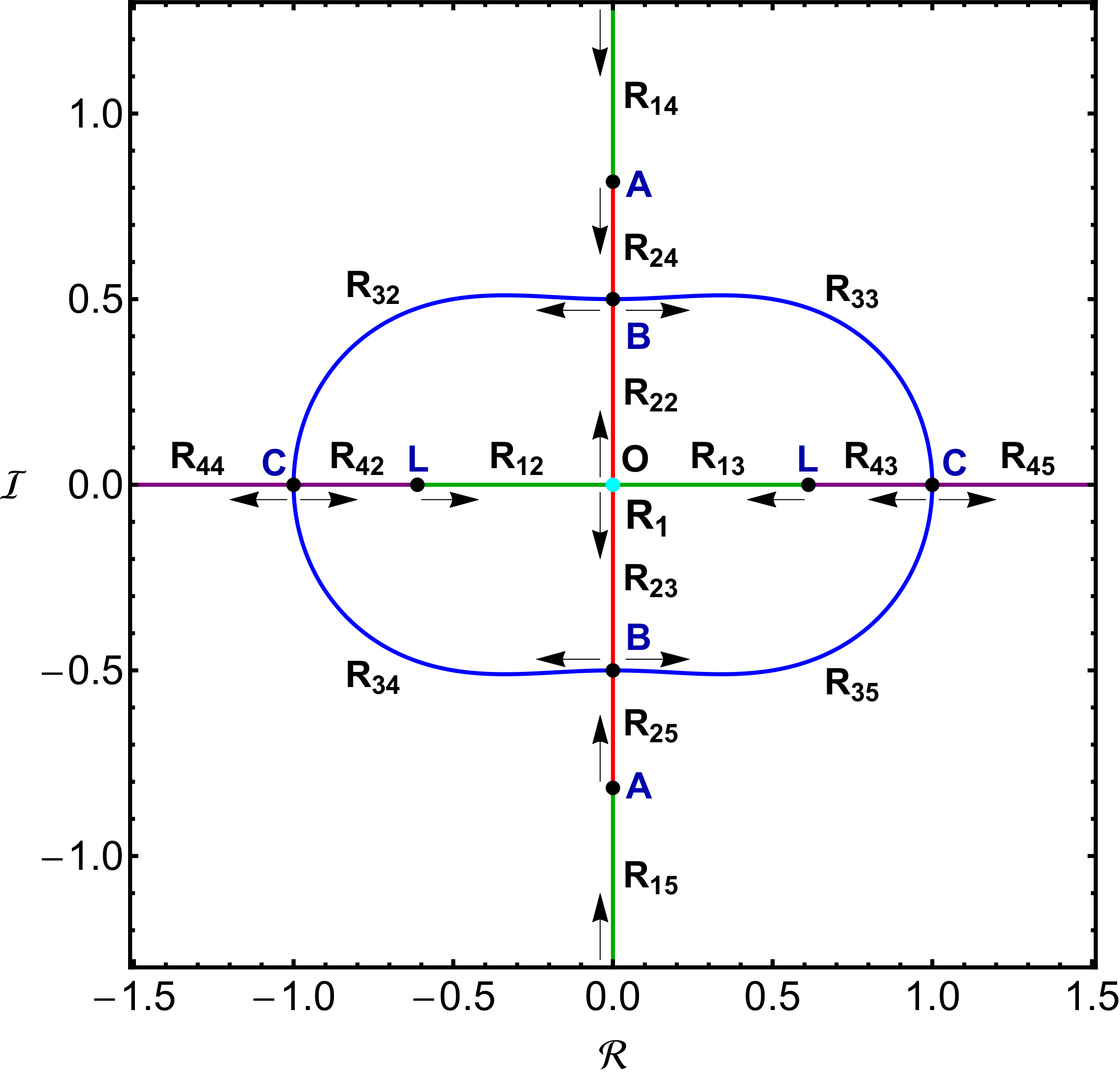}}
\caption{The space evolution of the four roots $R_{ij}$, $i,j = 1,...,4$ on the complex plane, when $A \in [-5,5]$. When $A < -1/18$ we have the roots $R_{12}$, $R_{13}$, $R_{14}$, and $R_{15}$ (green), when $A \in (-1/18, 0)$ we have the roots $R_{22}$, $R_{23}$, $R_{24}$, and $R_{25}$ (red), when $A \in (0, 5/6)$ we have the roots $R_{32}$, $R_{33}$, $R_{34}$, and $R_{35}$ (blue), while when $A > 5/6$ we have the roots $R_{42}$, $R_{43}$, $R_{44}$, and $R_{45}$ (purple). The arrows indicate the movement direction of the roots, as the value of the oblateness coefficient increases. The black dots (points A, B, and C) correspond to the three critical values of the oblateness coefficient $-1/18, 0, 5/6$, respectively, while the points L correspond to $A \to \pm \infty$. (Color figure online).}
\label{evol}
\end{figure}

It would be very interesting to determine how the positions of the four roots, on the complex plane, evolve as a function of the oblateness coefficient. Fig. \ref{evol} shows the parametric evolution of the four roots $R_{ij}$, $i,j = 1,...,4$, on the complex plane, when $A \in [-5,5]$, with $\mathcal{R} = Re[\mathz]$ and $\mathcal{I} = Im[\mathz]$. When $A \to - \infty$ the two real roots tend to $L = \pm \sqrt{3/2}/2$, while the two imaginary roots tend to infinity. As we proceed to higher values of $A$ all four roots tend to the central region. When $A = -1/18$ the two real roots collide at the origin which increases the multiplicity of the $z = 0$ root from 1 to 3. At the same time, the two imaginary roots are located at $A = \pm \sqrt{2/3}$ on the vertical axis. As soon as $A < -1/18$ a new pair of imaginary roots emerge from the origin $(0,0)$. As the value of $A$ increases approaching 0, all four imaginary roots move on collision courses. The collision occurs when $A = 0$, while the roots are exactly at $B = \pm 0.5$. For positive values of the oblateness coefficient (or in other words for oblate primaries) four complex roots emerge, one at each of the quadrants of the complex plane. As long as $A$ lies in the interval $(0, 5/6)$ the combined traces of the four complex roots create an oval shape. When $A = 5/6$ the four complex roots collide, in two pairs, on the horizontal axis, thus resulting to two real roots $C = \pm 1$ of multiplicity 2. For $A > 5/6$ two pairs of real roots emerge, while the roots of each pair move away from each other. Specifically, as $A \to \infty$ the outer roots $R_{44}$ and $R_{45}$ tend to infinity, while the roots $R_{42}$ and $R_{43}$ tend to $L = \pm \sqrt{3/2}/2$.

\section{The basins of attraction}
\label{bas}

A plethora of methods for numerically solving an equation with one variable parameter have been developed over the years. In this article we will consider and compare sixteen methods, whose order of convergence is varying from 2 to 16. In particular the methods under consideration are the following
\begin{enumerate}
  \item The Newton-Raphson's optimal method of second order \citep{CdB73}.
  \item The Halley's method of third order \citep{H64}.
  \item The Chebyshev's method of third order \citep{T64}
  \item The super Halley's method of fourth order \citep{GH01}.
  \item The modified super Halley's optimal method of fourth order \citep{CH08}.
  \item The King's method of fourth order \citep{K73}.
  \item The Jarratt's method of fourth order \citep{J66}.
  \item The Kung-Traub's optimal method of fourth order \citep{KT74}.
  \item The Maheshwari's optimal method of fourth order \citep{M09}.
  \item The Murakami's method of fifth order \citep{M78}.
  \item The Neta's method of sixth order \citep{N79}.
  \item The Chun-Neta's method of sixth order \citep{CN12}.
  \item The Neta-Johnson's method of eighth order \citep{NJ08}.
  \item The Neta-Petkovic's optimal method of eighth order \citep{NP10}.
  \item The Neta's method of fourteenth order \citep{N81}.
  \item The Neta's method of sixteenth order \citep{N81}.
\end{enumerate}
The analytical expressions of all the above-mentioned iterative schemes are given in the Appendix of \cite{Z17b}.

All the computational methodology that we are going to use in order to classify the initial conditions on the complex plane are described in detail in Section 3 of \citet{Z17b}.

\begin{figure*}[!t]
\centering
\resizebox{\hsize}{!}{\includegraphics{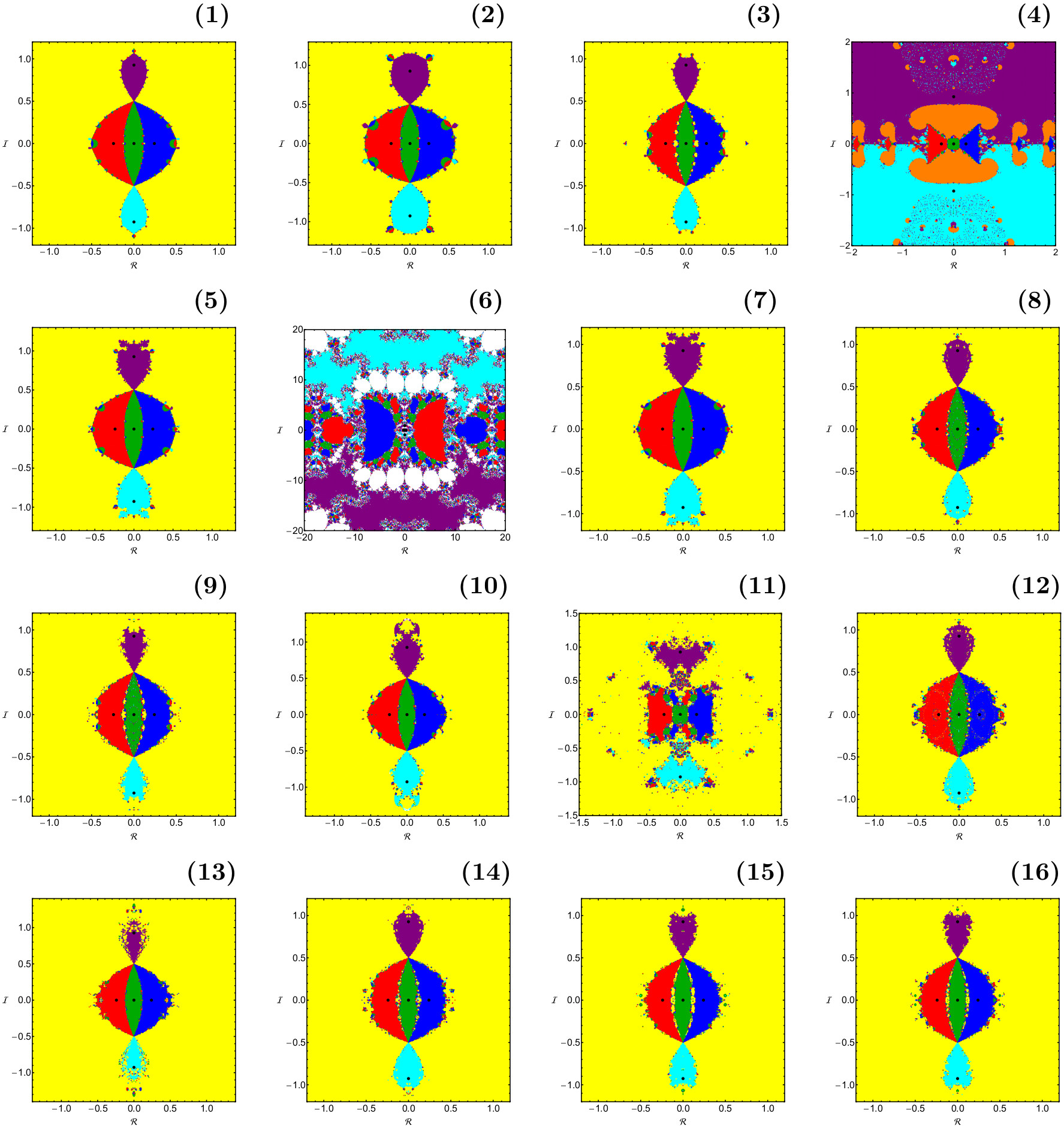}}
\caption{Basin diagrams on the complex plane, when $A = -0.1$. Black dots are used for indicating the position of the five roots. The color code is as follows: $R_1$ root (green); $R_2$ root (red); $R_3$ root (blue); $R_4$ root (purple); $R_5$ root (cyan); false convergence to $\pm 0.5i$ (orange); convergence to infinity (yellow); non-converging points (white). The numbers of the panels correspond to the numerical methods, as they have been listed at the beginning of Section \ref{bas}. (Color figure online).}
\label{c1}
\end{figure*}

\begin{figure*}[!t]
\centering
\resizebox{\hsize}{!}{\includegraphics{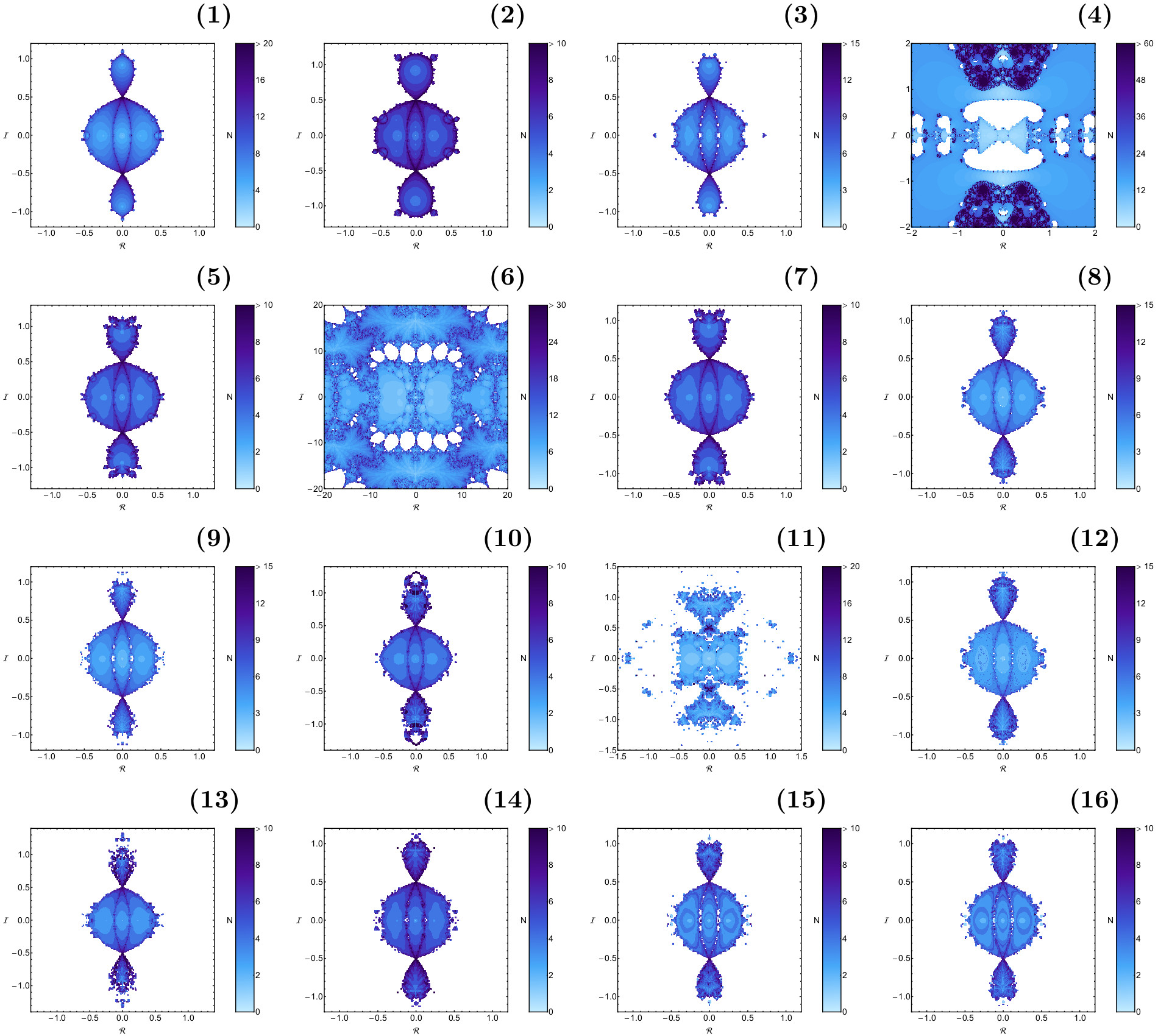}}
\caption{Color-coded diagrams showing the distributions of the required number $N$ of iterations for the corresponding basins of convergence of Fig. \ref{c1}. White color is used for all the ill-behaved (False converging and non-converging) initial conditions. (Color figure online).}
\label{n1}
\end{figure*}

\begin{figure*}[!t]
\centering
\resizebox{\hsize}{!}{\includegraphics{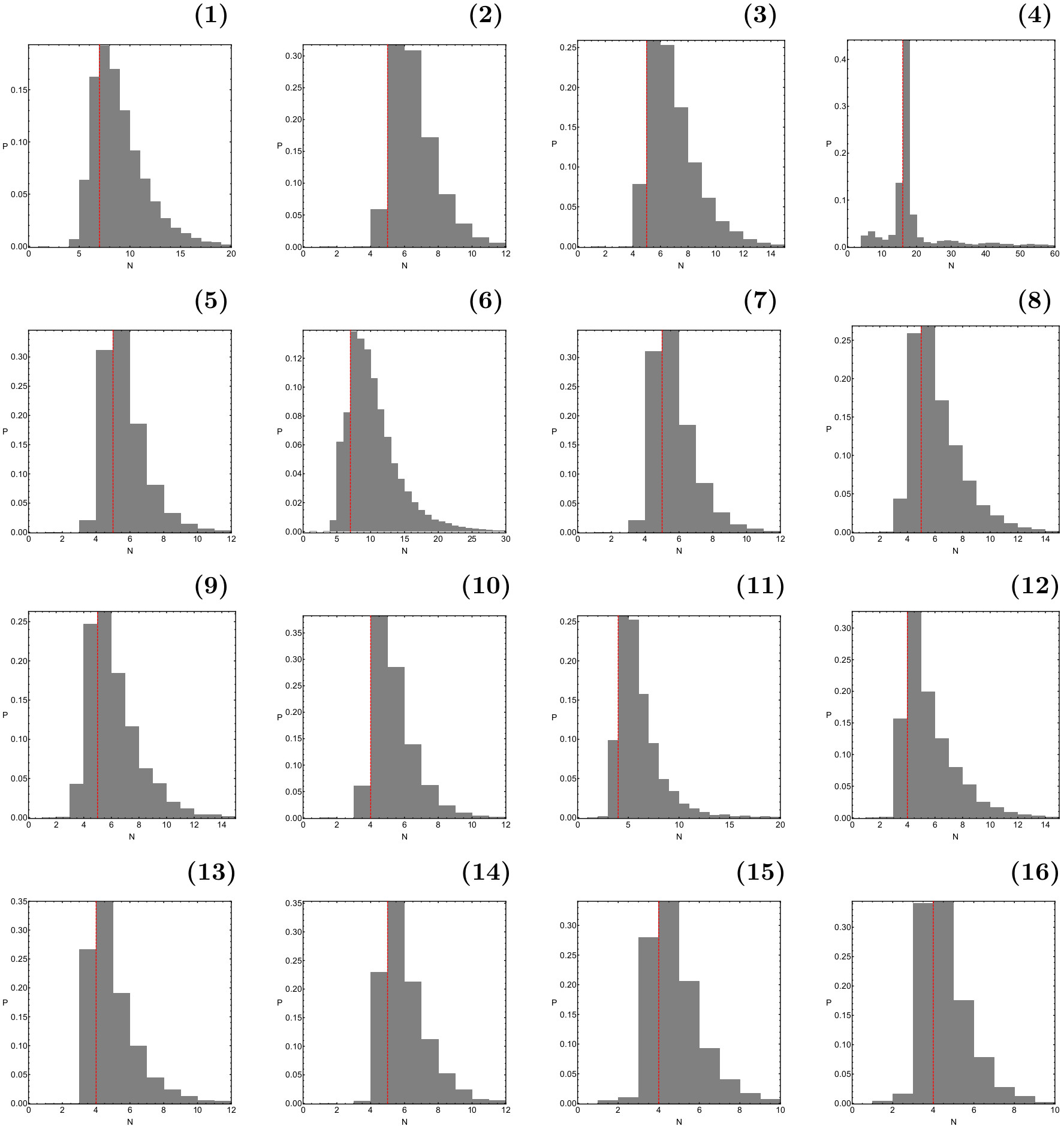}}
\caption{Histograms depicting the probability distributions of the required number $N$ of iterations for the corresponding basins of convergence of Fig. \ref{c1}. The position of the most probable number $N^{*}$ of iterations is indicated using a dashed vertical red line. (Color figure online).}
\label{p1}
\end{figure*}

\begin{figure*}[!t]
\centering
\resizebox{\hsize}{!}{\includegraphics{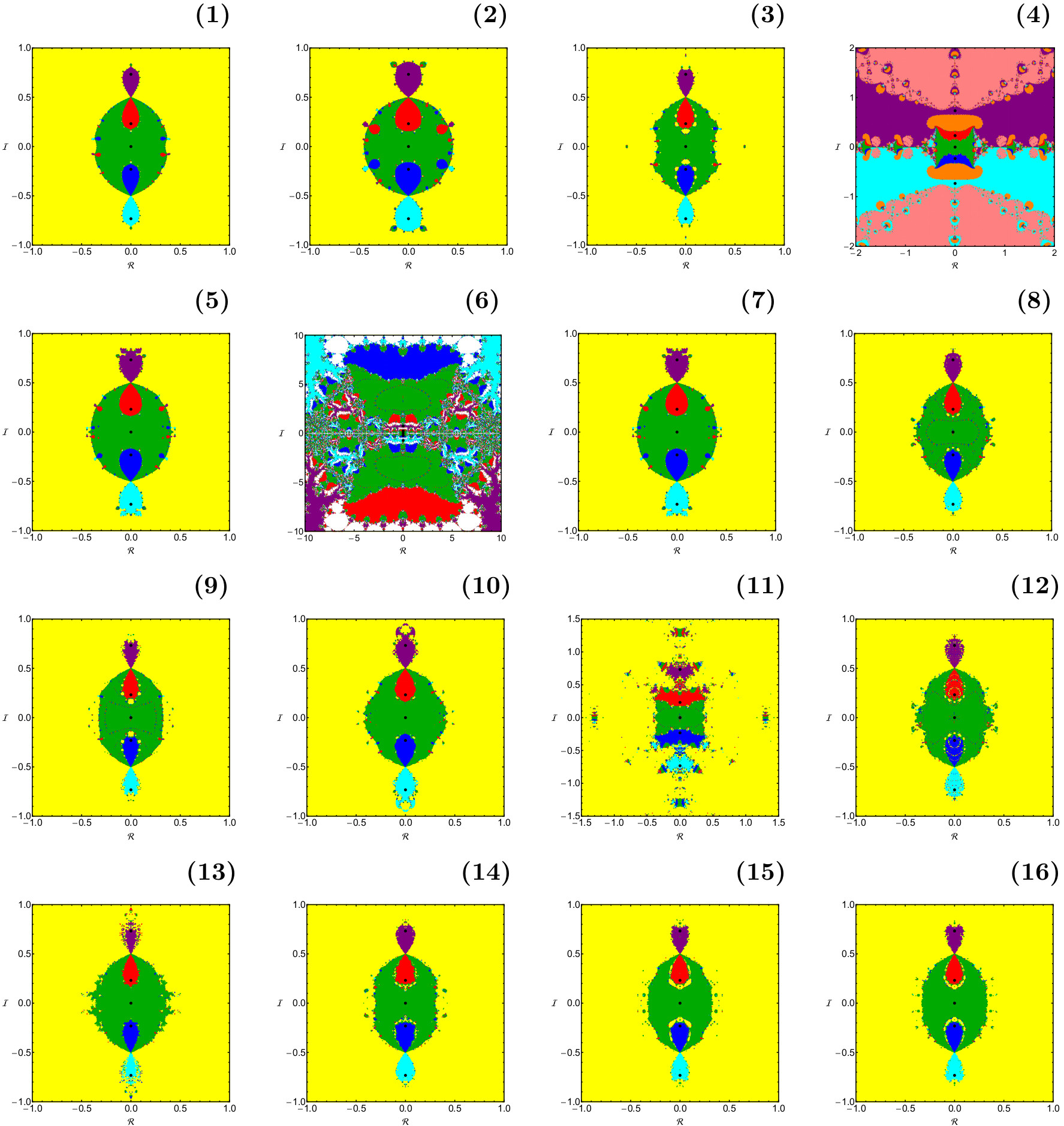}}
\caption{Basin diagrams on the complex plane, when $A = -0.1$., when $A = -0.03$. Black dots are used for indicating the position of the five roots. The color code is as follows: $R_1$ root (green); $R_2$ root (red); $R_3$ root (blue); $R_4$ root (purple); $R_5$ root (cyan); false convergence to $\pm 0.5i$ (orange); convergence to infinity (yellow); non-converging points (white). The numbers of the panels correspond to the numerical methods, as they have been listed at the beginning of Section \ref{bas}. (Color figure online).}
\label{c2}
\end{figure*}

\begin{figure*}[!t]
\centering
\resizebox{\hsize}{!}{\includegraphics{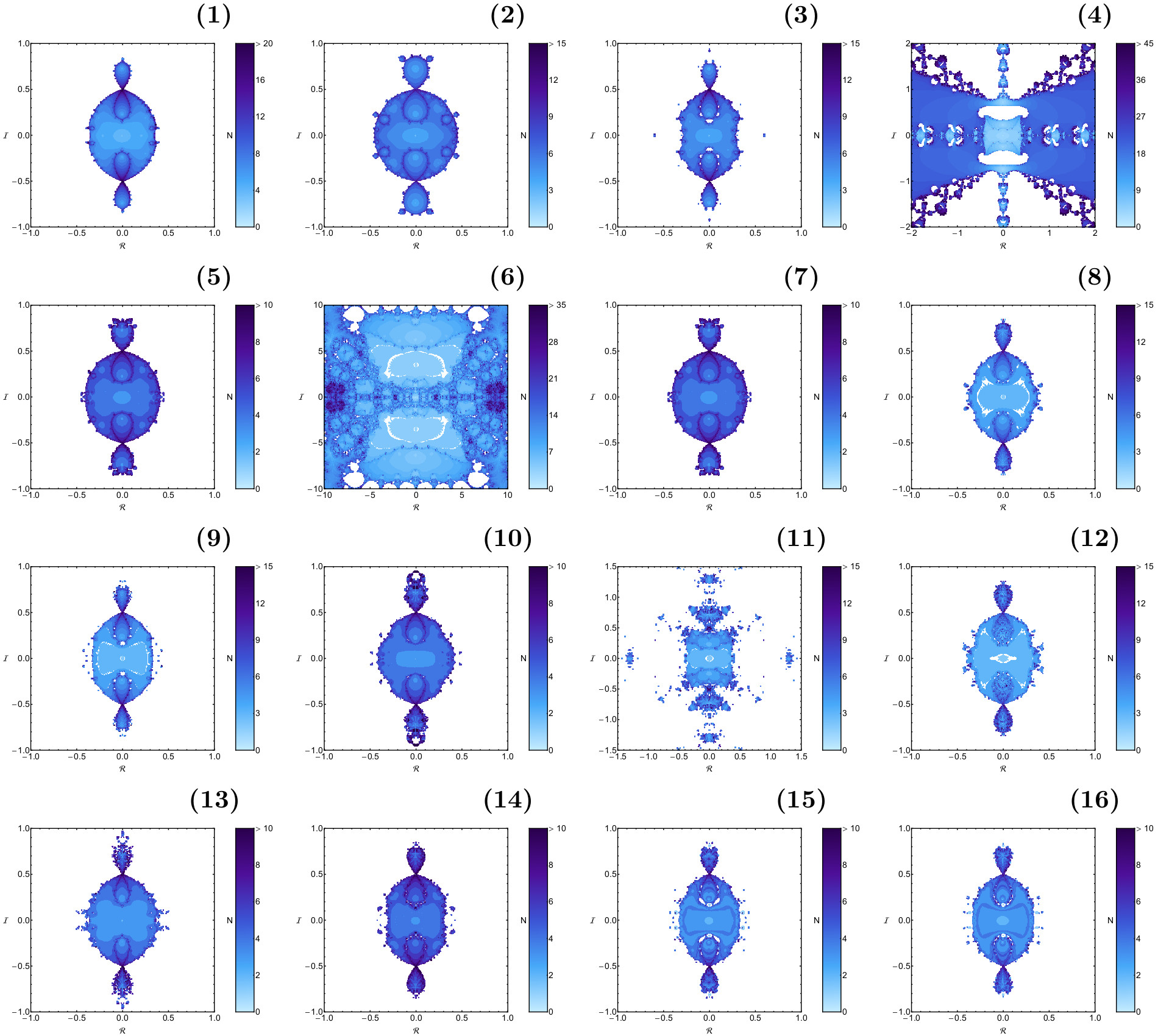}}
\caption{Color-coded diagrams showing the distributions of the required number $N$ of iterations for the corresponding basins of convergence of Fig. \ref{c2}. White color is used for all the ill-behaved (False converging and non-converging) initial conditions. (Color figure online).}
\label{n2}
\end{figure*}

\begin{figure*}[!t]
\centering
\resizebox{\hsize}{!}{\includegraphics{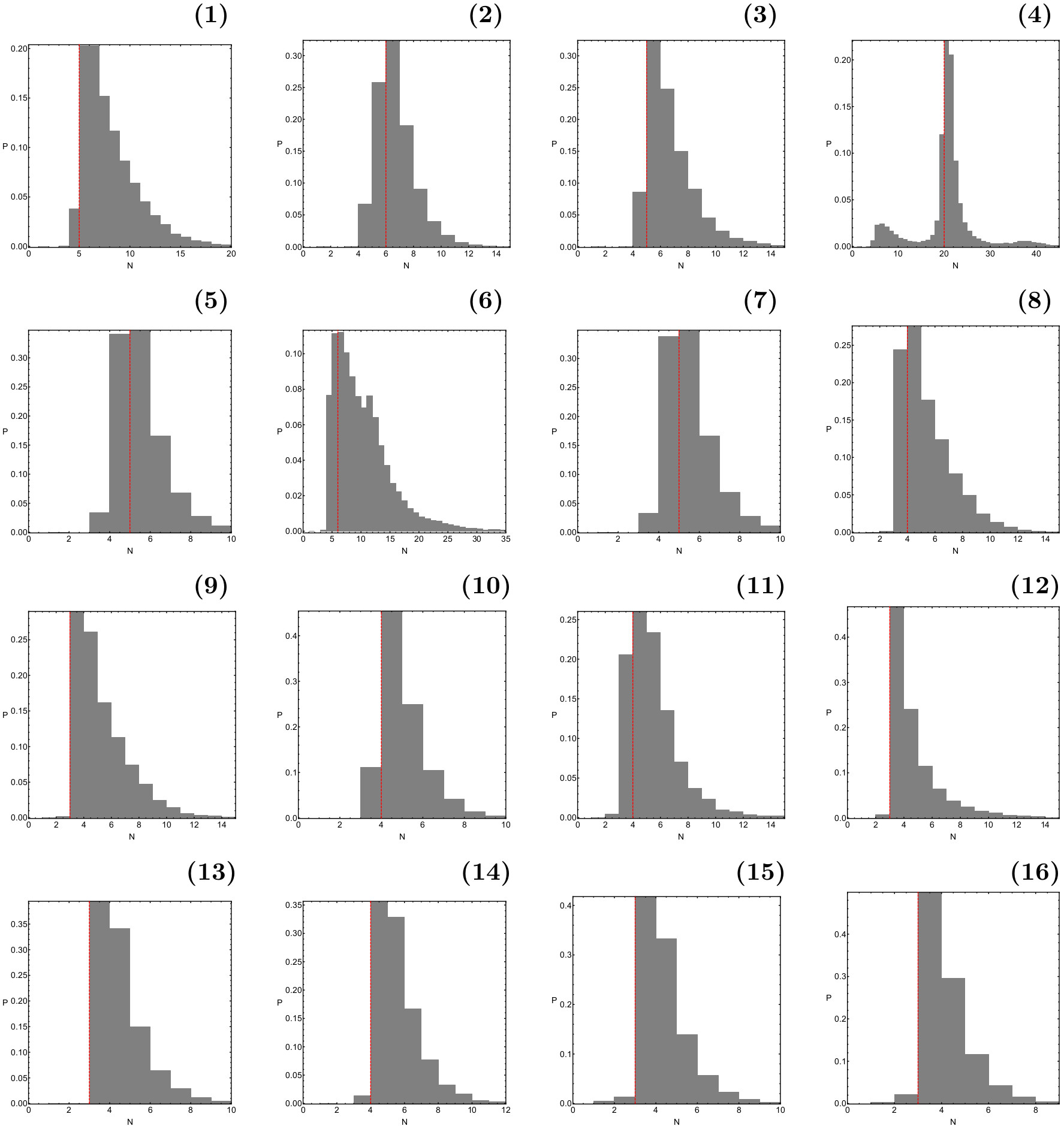}}
\caption{Histograms depicting the probability distributions of the required number $N$ of iterations for the corresponding basins of convergence of Fig. \ref{c2}. The position of the most probable number $N^{*}$ of iterations is indicated using a dashed vertical red line. (Color figure online).}
\label{p2}
\end{figure*}

\begin{figure*}[!t]
\centering
\resizebox{\hsize}{!}{\includegraphics{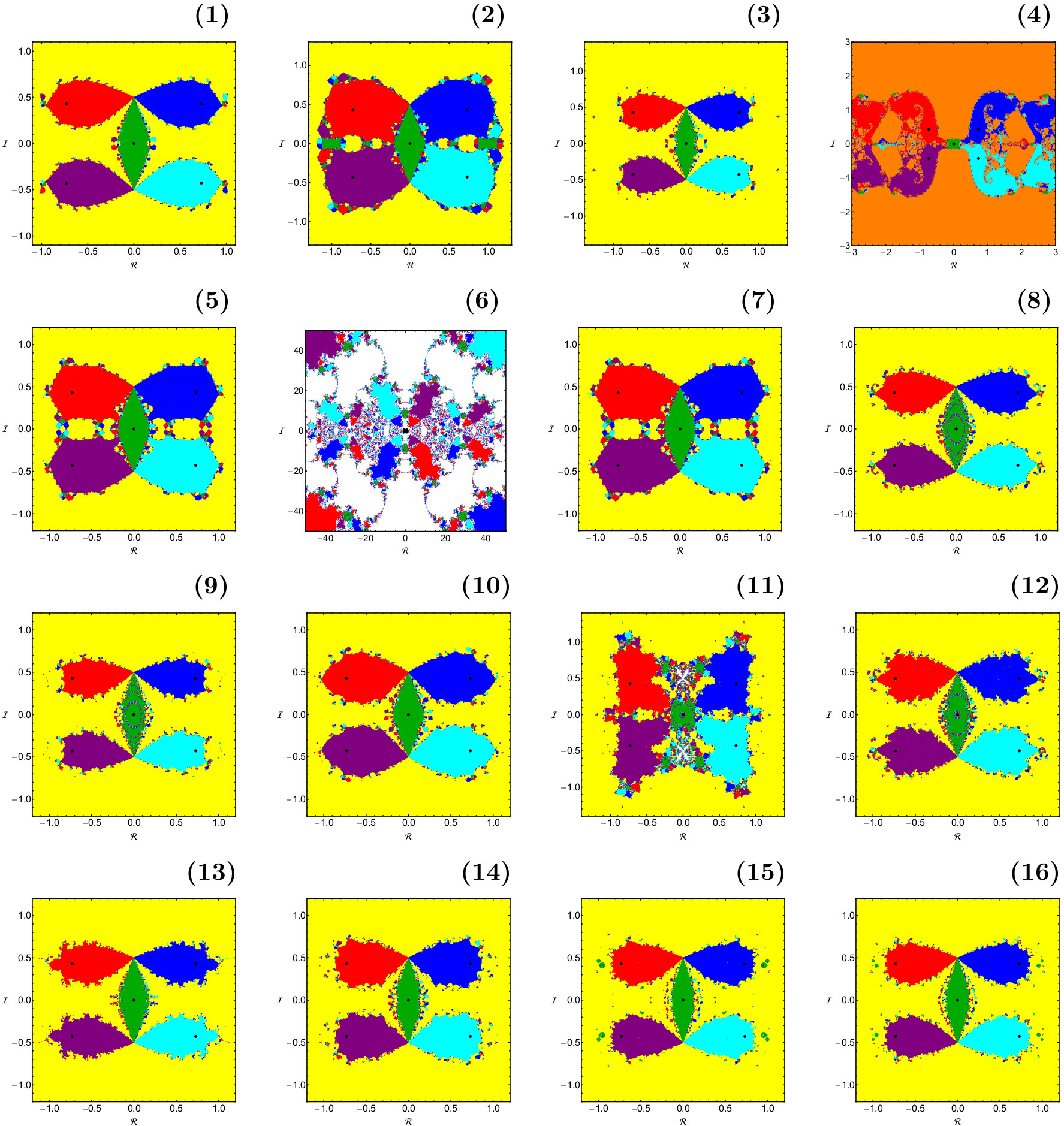}}
\caption{Basin diagrams on the complex plane, when $A = 0.4$. Black dots are used for indicating the position of the five roots. The color code is as follows: $R_1$ root (green); $R_2$ root (red); $R_3$ root (blue); $R_4$ root (purple); $R_5$ root (cyan); false convergence to $\pm 0.5i$ (orange); convergence to infinity (yellow); non-converging points (white). The numbers of the panels correspond to the numerical methods, as they have been listed at the beginning of Section \ref{bas}. (Color figure online).}
\label{c3}
\end{figure*}

\begin{figure*}[!t]
\centering
\resizebox{\hsize}{!}{\includegraphics{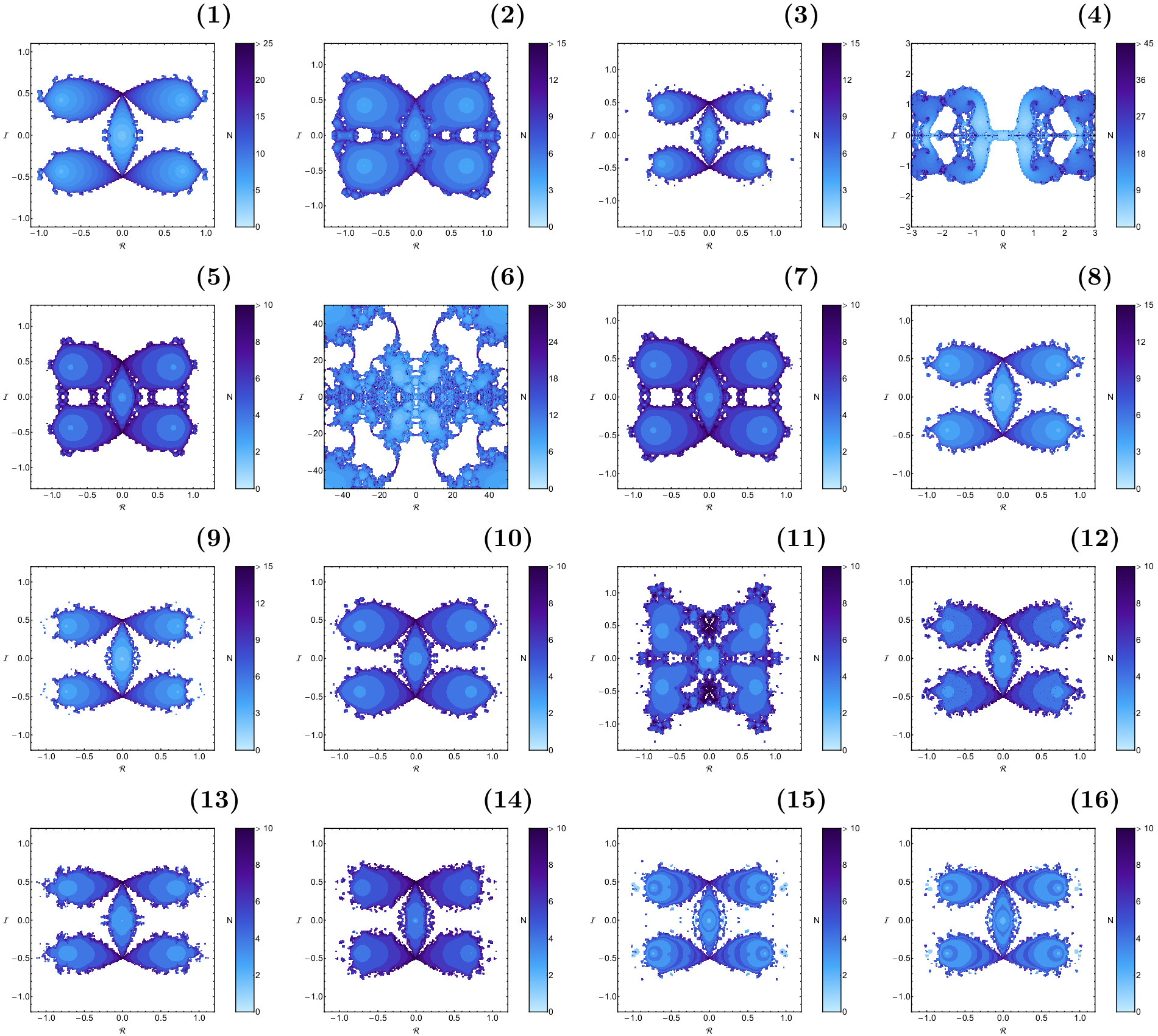}}
\caption{Color-coded diagrams showing the distributions of the required number $N$ of iterations for the corresponding basins of convergence of Fig. \ref{c3}. White color is used for all the ill-behaved (False converging and non-converging) initial conditions. (Color figure online).}
\label{n3}
\end{figure*}

\begin{figure*}[!t]
\centering
\resizebox{\hsize}{!}{\includegraphics{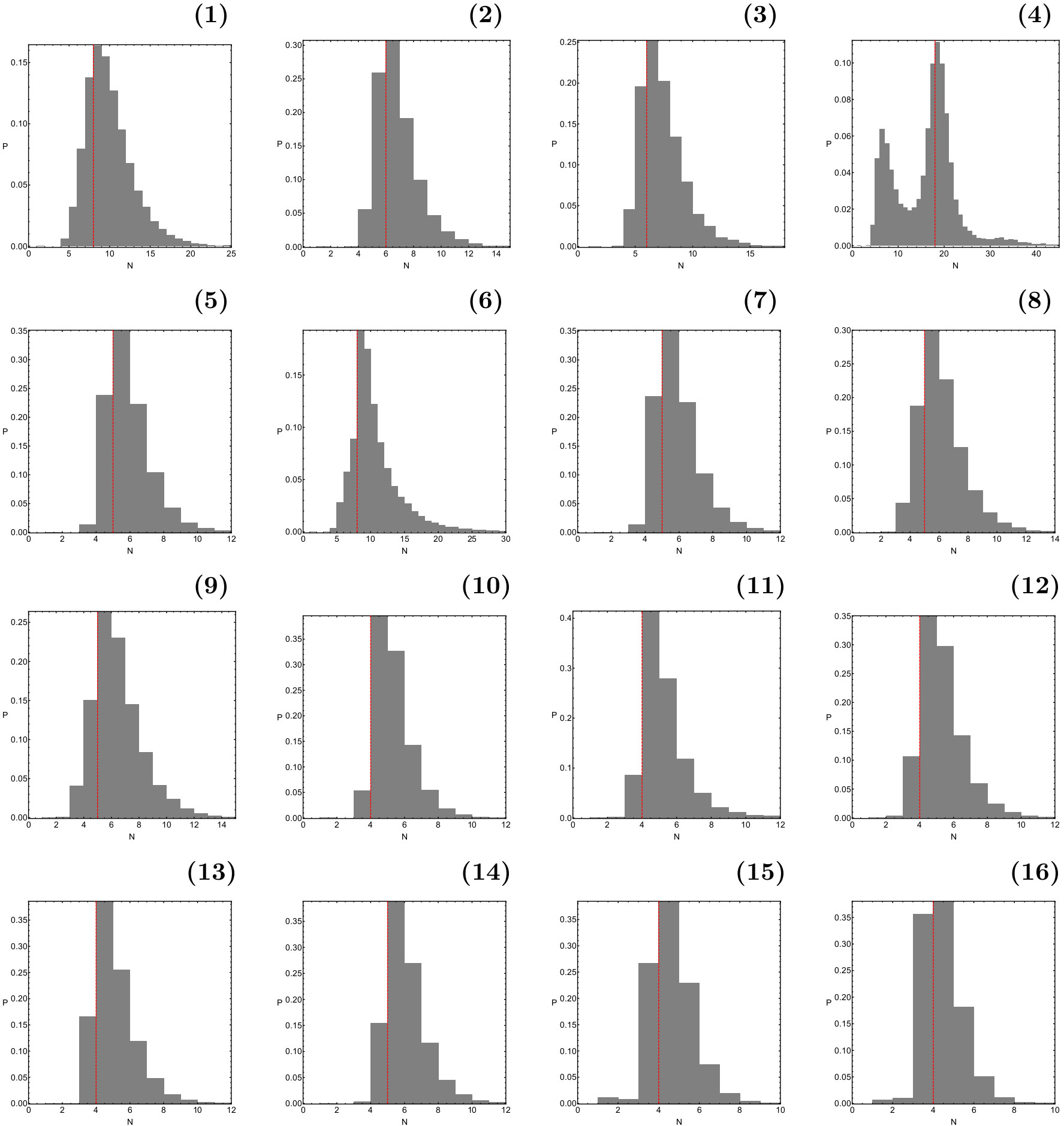}}
\caption{Histograms depicting the probability distributions of the required number $N$ of iterations for the corresponding basins of convergence of Fig. \ref{c3}. The position of the most probable number $N^{*}$ of iterations is indicated using a dashed vertical red line. (Color figure online).}
\label{p3}
\end{figure*}

\begin{figure*}[!t]
\centering
\resizebox{\hsize}{!}{\includegraphics{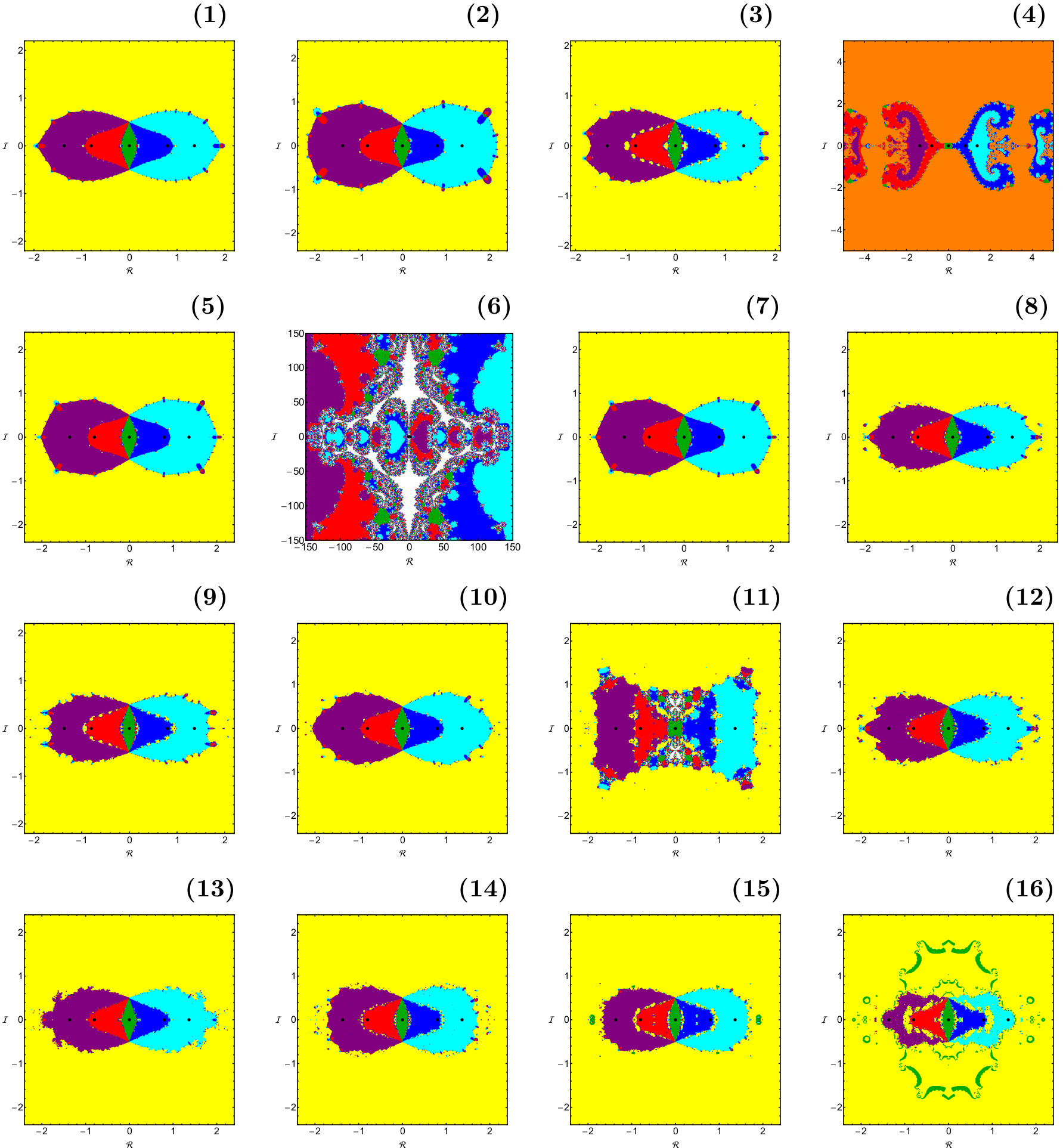}}
\caption{Basin diagrams on the complex plane, when $A = 1$. Black dots are used for indicating the position of the five roots. The color code is as follows: $R_1$ root (green); $R_2$ root (red); $R_3$ root (blue); $R_4$ root (purple); $R_5$ root (cyan); false convergence to $\pm 0.5i$ (orange); convergence to infinity (yellow); non-converging points (white). The numbers of the panels correspond to the numerical methods, as they have been listed at the beginning of Section \ref{bas}. (Color figure online).}
\label{c4}
\end{figure*}

\begin{figure*}[!t]
\centering
\resizebox{\hsize}{!}{\includegraphics{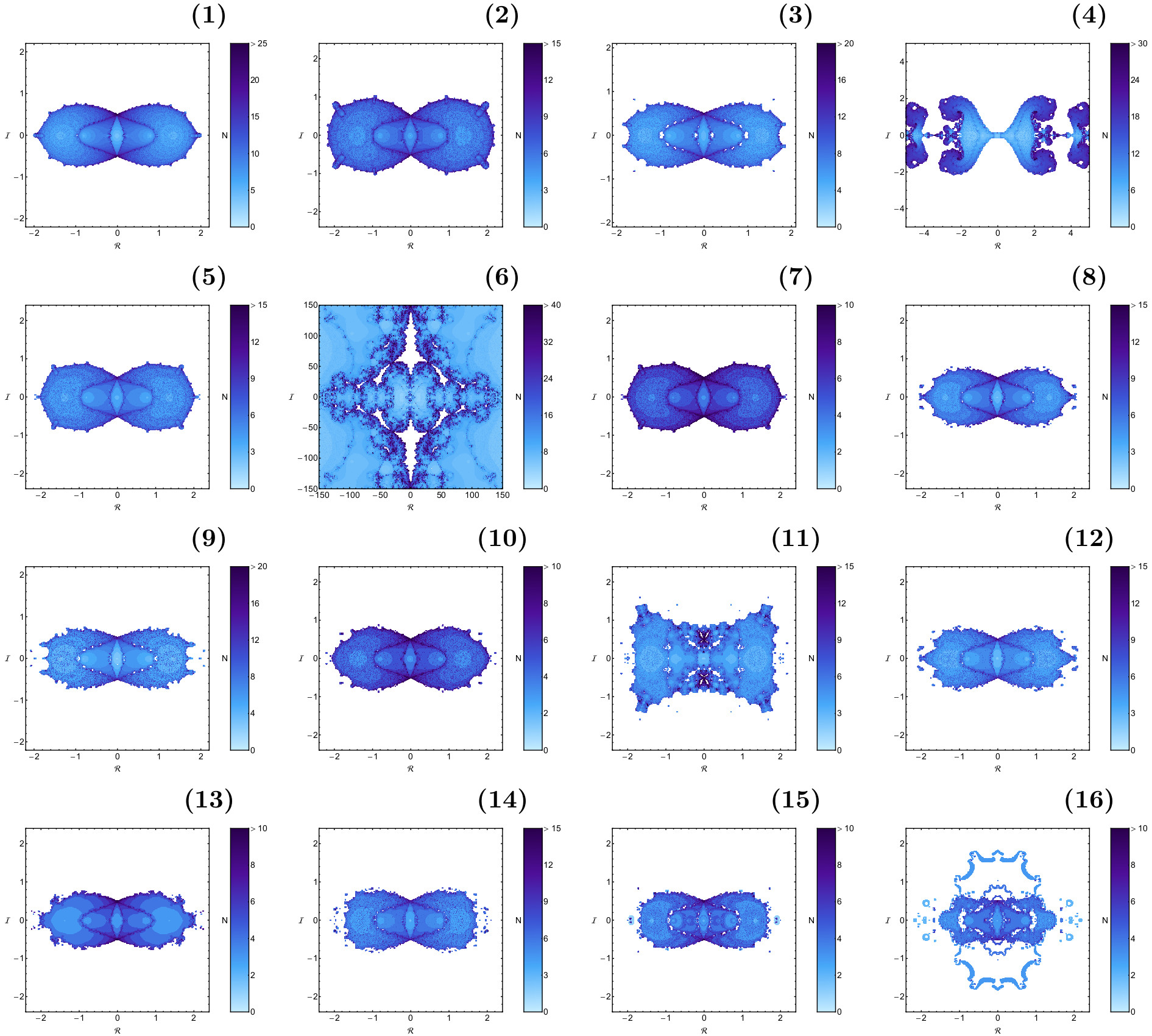}}
\caption{Color-coded diagrams showing the distributions of the required number $N$ of iterations for the corresponding basins of convergence of Fig. \ref{c4}. White color is used for all the ill-behaved (False converging and non-converging) initial conditions. (Color figure online).}
\label{n4}
\end{figure*}

\begin{figure*}[!t]
\centering
\resizebox{\hsize}{!}{\includegraphics{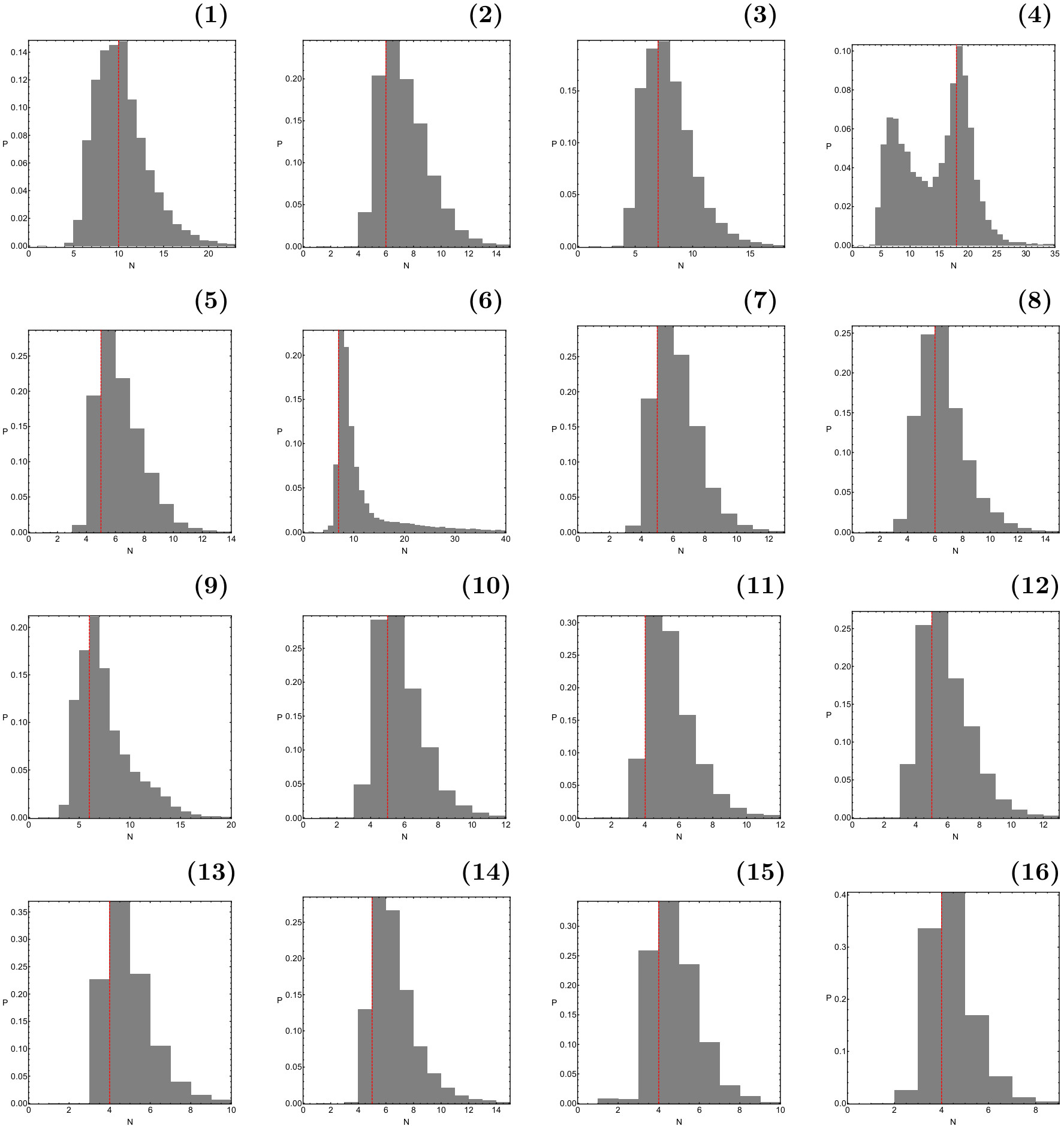}}
\caption{Histograms depicting the probability distributions of the required number $N$ of iterations for the corresponding basins of convergence of Fig. \ref{c4}. The position of the most probable number $N^{*}$ of iterations is indicated using a dashed vertical red line. (Color figure online).}
\label{p4}
\end{figure*}

\subsection{Case I: Two real and two imaginary roots, along with (0,0)}
\label{ss1}

We begin with the first case, that is when $A < -1/18$, where there are two real and two imaginary roots, along with the $(0,0)$ root. The basins of attraction on the complex plane, for $A = -0.1$, using the sixteen iterative schemes are presented in Fig. \ref{c1}. Our calculations suggest that the convergence structure on the complex plane for the majority of the numerical methods is, in general terms, very similar. In particular, for all the numerical methods except for the super Halley and the King methods we observe the following aspects:
\begin{itemize}
  \item The extent of all the basins of attractions, associated with the five roots, is finite. Furthermore, the shape of all the attracting regions resemble the shape of a lobe.
  \item Around the attracting domains the complex plane is covered by a unified sea of initial conditions (yellow regions) which do not converge to any of the five roots of the system. In fact for these initial conditions all the iterative schemes (except for the super Halley and the King methods) lead, sooner or later, to extremely large numbers, which is a numerical indication that these particular initial conditions lead to infinity.
  \item It is interesting to note that the convergence structure, corresponding to Neta iterative scheme, is a bit different, with respect to the convergence structures of the other thirteen methods.
\end{itemize}

On the other hand, the convergence structures of both the super Halley and the King methods have significant differences comparing to the rest of the numerical methods. In panel (4) of Fig. \ref{c1} it is seen that all the basins of attraction, corresponding to the five roots, extend to infinity. Apart from the usual basins of convergence there exist additional basins (orange regions) in which the initial conditions display a strange behavior. Our analysis indicates that the initial conditions which form these basins converge to $\pm 0.5i$, which are not roots of the system. This means that the super Halley method for a considerable amount of initial conditions exhibits a false convergence.

According to panel (6) of Fig. \ref{c1} all the basins of attraction also extend to infinity in the case of the King method. Between the convergence regions we can identify several areas (white) in which the initial conditions do not converge to any of the five roots of the system. Initially we suspected that maybe these initial conditions are just extremely slow converging nodes. To check this we increased the maximum allowed number of iterations from 500 to 50000 and we reclassified these initial conditions. We found that for all these initial conditions, during the iterative procedure, the final state smoothly oscillates between two complex numbers of the form $\pm a + bi$. Therefore, we argue that for these initial conditions we have strong numerical evidence that they do not converge to any of the roots of the system.

Another interesting aspect, shown in the convergence diagram of the King method, concerns the geometry of the CCD. More precisely, one can observe that the overall pattern, especially in the vicinity of the basin boundaries, is very noisy or in other words highly fractal. This directly implies that for the initial conditions inside these chaotic domains it is extremely difficult (or even impossible) to know beforehand their final state (root). At this point, it should be noted that when we state that an area is fractal we simply mean that it displays a fractal-like geometry, without using any quantitative arguments, such as the fractal dimension, as in \citet{AVS01,AVS09}.

In Fig. \ref{n1} we provide, using tones of blue, the corresponding distributions of the number $N$ of the required iterations for obtaining the basins of attraction shown in Fig. \ref{c1}. In almost all cases we observe the expected behavior according to which the fastest converging nodes are those with initial conditions inside the basins of attraction, while the slowest initial conditions are those located in the vicinity of the basin boundaries. However in panel (4) it becomes evident that a considerable amount of initial conditions require more than 50 iterations for converging to one of the two imaginary roots $R_4$ and $R_5$, while in all other cases, the vast majority of the initial conditions converge within the first 20 iterations. We suspect that the phenomenon of the extremely slow converging points is directly related with the existence of false converging points. Moreover, it should be emphasized that the regions on the complex plane, in which the extremely slow converging nodes are located, are highly fractal.

The corresponding probability distributions of iterations are given in Fig. \ref{p1}. The definition of the probability $P$ is the following: assume that after $N$ iterations $N_0$ initial conditions on the complex plane converge to one of the roots of the system. Then $P = N_0/N_t$, where $N_t$ is the total number of initial conditions in every CCD. Our results suggest that for almost all the numerical methods more than 98\% of the initial conditions converge within the first 30 iterations, while only for the super Halley method the tail of the corresponding histogram extends to 60 iterations. The most probable number of iterations $N^{*}$ (see the vertical, dashed, red lines in the histograms) seems, in general terms, to decrease as we proceed to numerical methods of higher order.

\subsection{Case II: Four imaginary roots, along with (0,0)}
\label{ss2}

In this case, where $-1/18 < A < 0$ the system admits four imaginary roots, along with the classical $(0,0)$ root. In Fig. \ref{c2} we provide the CCDs for the sixteen numerical methods, when $A = -0.03$. Once more, the convergence properties of most of the iterative schemes are very similar, while the only two cases which display complete different patterns are those corresponding to super Halley and King methods. For both these methods the structure of the corresponding CCDs is highly complex.

For the super Halley method it is seen in panel (4) of Fig. \ref{c2} that apart from the basins of attraction, associated with the five roots, there exist two additional types of basins. The first type corresponds to initial conditions which display a false convergence to $\pm 0.5i$ (orange regions), while the second type corresponds to initial conditions that exhibit a false convergence to complex numbers of the form $\pm a \pm bi$ (pink regions). In the case of the King method (see panel (6) of Fig. \ref{c2})we have again the appearance of non-converging initial conditions which infinitely oscillate between two complex numbers of the form $\pm a + bi$.

The corresponding distributions of the number of iterations and the probability are given in Figs. \ref{n2} and \ref{p2}, respectively. It should be mentioned that the most smooth distributions of the probability histograms correspond to numerical methods where either the initial conditions converge to one of the five roots or tend to infinity. The most noisy histograms on the other hand, in which multiple peaks are present (see e.g., panel (4) of Fig. \ref{p2}) correspond to problematic numerical methods in which the phenomenon of false convergence occurs.

\subsection{Case III: Four complex roots, along with (0,0)}
\label{ss3}

The next case under consideration corresponds to $0 < A < 5/6$, when there are four complex roots, along with the central root $(0,0)$. The basins of attraction on the complex plane for the sixteen methods, when $A = 0.4$ are presented in Fig. \ref{c3}. It is seen that only the CCDs corresponding to super Halley and King methods are different, while almost all the other CCDs have similar convergence patterns.

For the super Halley method, shown in panel (4) of Fig. \ref{c3}, we observe that the vast majority of the complex plane (orange regions) is covered by initial conditions for which the super Halley iterative method displays a false convergence to $\pm 0.5i$. In panel (6), regarding the King method, we encounter again a substantial amount of non-converging initial conditions for which the iterative scheme oscillates between two imaginary roots. Looking carefully at panel (11) of Fig. \ref{c3} one can also identify a small portion of non-converging initial conditions in the case of the Neta method. Our calculations indicate that for these nodes the iterative scheme oscillates between two complex numbers, which do not coincide with the complex roots of the system.

In Fig. \ref{n3} we provide the corresponding distributions of the required number $N$ of iterations, while the corresponding probability distributions are shown in Fig. \ref{p3}. Combining the results of both these figures we may conclude that the highest numbers of required iterations $(N > 25)$ are observed in numerical methods with problematic behavior, where false and non-converging points are present.

\subsection{Case IV: Four real roots, along with (0,0)}
\label{ss4}

We close our numerical investigation with the last case, that is for $A > 5/6$, when the system admits four real roots, along with the universal root $\mathz = 0$. Fig. \ref{c4} depicts the basins of attraction of the sixteen numerical methods, when $A = 1$. As in the previous subsections, the convergence properties of most of the examined numerical methods are, in general terms, very similar.

Once more, for the super Halley method (see panel (4) of Fig. \ref{c4}) we encountered the phenomenon of initial conditions with false convergence to $\pm 0.5i$. In the same vein, the phenomenon of non-converging initial conditions is also observed for the King method (see panel (6)). At this point we should note the complicated basin structures (with the highly fractal basin boundaries) on the complex plane which are produced by the King numerical method. Non-converging initial conditions are also present in the case of the Neta method (see panel (11) of Fig. \ref{c4}). However in this case the corresponding iterative procedure oscillates between two complex numbers, while in the King method there is an infinite oscillation between two imaginary roots.

In Fig. \ref{n4} one can observe how the corresponding numbers $N$ of the required iteration are distributed on the complex plane, for the numerical methods presented in Fig. \ref{c4}. In panel (6) of Fig. \ref{n4} it is clearly seen how the extremely slow converging points (when $N > 30$) are located in the vicinity of the fractal basin boundaries. Indeed, in panels (4) and (6) of Fig. \ref{p4} we see how the tails of the corresponding histograms are much more extended with respect to histograms of all the other numerical methods.

Before ending this section we would like to state that mainly for saving space we did not present any results regarding the three critical values of the oblateness coefficient $A = (-1/18, 0, 5/6)$. In fact, for these particular values of $A$ the system has either one or three roots and therefore the overall structure of the basins of attraction is less interesting.

\section{Concluding remarks}
\label{conc}

In this work we used a large variety of numerical methods in order to reveal the basins of attraction on the complex plane in the circular Sitnikov problem, when the two primary bodies are either prolate or oblate spheroids. All the magnificent basin structures on the complex plane were identified by classifying dense grids of initial conditions, using the corresponding iterative schemes. In particular, we managed to determine how the geometry of the convergence structures changes as a function of the order of the applied numerical methods. Furthermore, all the correlations between the attracting domains and the corresponding distributions of the probability as well as the required number of iterations have been successfully established.

It should be emphasized that this is the first time that the basins of attraction in the circular Sitnikov problem with spheroid primaries are numerically investigates in such a thorough and systematic manner, using a plethora of numerical methods. On this basis, we argue that all the presented numerical results are novel, while they add considerably to our existing knowledge on the field of basins of attraction.

The following list contains the most important conclusions of our numerical analysis:
\begin{enumerate}
  \item For almost all the numerical methods (except for the super Halley and the King methods) and for all the examined cases (regarding the nature of the five roots of the system) all basins of attractions are finite. On the other hand, for the super Halley and the King methods all the basins of convergence extend to infinity.
  \item For all the methods where the basins of attractions are finite, it was found that the rest of the complex plane is covered by initial conditions for which the corresponding iterative schemes lead to extremely large numbers (in other words, we have a numerical indication that these nodes lead to infinity).
  \item We observed that not all numerical methods display the same degree of efficiency. In particular, for the super Halley method we identified a non-zero amount of problematic initial conditions which display false convergence to final states which are different from the roots of the system.
  \item Our experiments indicated that for the King and the Neta methods there exist basins of initial conditions which do not converge not even after 50000 iterations. Additional numerical computations revealed that for these initial conditions the corresponding iterative schemes infinitely oscillate between two (imaginary or complex) numbers, which directly implies that these nodes are true non-converging points.
  \item The most complicated convergence structures on the complex plane, full of highly fractal basin boundaries, correspond to iterative methods with problematic behavior, that is when false or non-converging points are present.
  \item For the majority of the numerical methods more than 98\% of the classified initial conditions converge, to one of the five roots, within the first 25 iterations. Only for the problematic methods (super Halley and King) the required number $N$ of iterations extends to more than 30 iterations.
  \item If we exclude the two most problematic methods (super Halley and King) then we may conclude that there is a clear relation between the convergence speed of the iterative schemes and the order of the methods. More specifically, the most probable number of iterations $N^{*}$ seems to decrease, with increasing order of the method.
\end{enumerate}

A double precision numerical routine, written in standard \verb!FORTRAN 77! \citep{PTVF92}, was used for the classification of the initial conditions on the complex plane. Using a Quad-Core i7 2.4 GHz PC we needed about 4 minutes of CPU time, for performing the classification in each grid of initial conditions. Moreover, all the graphical illustration of the paper has been created using the latest version 11.3 of Mathematica$^{\circledR}$ \citep{W03}.

We hope that the presented numerical outcomes, regarding the convergence properties of the circular Sitnikov problem with spheroid primaries, to be useful in the active field of numerical methods and the associated basins of attraction.

\section*{Acknowledgments}
\footnotesize

I would like to express my warmest thanks to the two anonymous referees for the careful reading of the manuscript and for all the apt suggestions and comments which allowed us to improve both the quality and the clarity of the paper.

\end{document}